# Infrared thermal imaging camera to measure low temperature thermal fields

E. Gordiyenko[1], Yu. Fomenko[1,a], G. Shustakova[1], G. Kovalov[2], S. Shevchenko[1]

[1]*B.Verkin Institute for Low Temperature Physics and Engineering, NAS of Ukraine, Nauky ave. 47, 61103 Kharkiv, Ukraine*

[2]*Institute for Problems of Cryobiology and Cryomedicine, NAS of Ukraine, Pereyaslavska str. 23, 61016 Kharkiv, Ukraine*

To measure low-temperature thermal fields, we have developed a single-element cooled thermal imaging camera for a spectral range of 8÷14 µm with an internal shutter for radiometric calibration. To improve the accuracy of measuring the temperature of cold objects, we used a shutter with a combined emissivity as an internal reference source of radiation at the input of the device optical unit. With this aim a small mirror was fixed in the center on its surface covered black, thereby ensuring an efficient reflection of radiation in a wide spectral range of wavelengths. When processing the signal for each pixel of the thermal image, the differential value of the detector response to the shutter blackened and mirror areas was used as a reference. A relative measurement error of 3% was obtained for the studied objects with a temperature of −150 °C. The device was successfully used for remote study of thermal field dynamics during freeze-thawing of biological tissues *in vivo*.

## I. INTRODUCTION

Most of the tasks of modern thermal diagnostics require not only a qualitative display of thermal fields, but also their quantitative assessment with a minimum error [1]. Of great scientific and practical interest is the remote measurement of the dynamics of low-temperature thermal fields, in particular, on the surface of biological objects during cryodestruction of the tissues with pathologies. When using a cryo-instrument cooled with liquid nitrogen, the temperature of the tissue can reach −150 °C and below. Of particular interest is the task of remote non-invasive monitoring of the movement of the boundary of the zone of tissue thermal damage during cryotherapy. The temperatures required for the guaranteed destruction of malignant tumors reach −40…−60 °C [2, 3].

The low-temperature limit of the range in which a thermal imager can make measurements is determined, first of all, by the sensitivity of the detector, the spectral range of sensitivity, and the parameters of the optics used. Modern detectors are sensitive enough to detect weak radiation from cold objects. The main problem is to ensure acceptable accuracy of low-temperature measurements. The heat flux from the surface of cold objects is so small that it becomes comparable to the heat flux from the surface of the elements of the optical path of the device. This situation seriously reduces low-temperature measurement limit unless special methods and approaches are used in the device's hardware and software, which eliminate the influence of thermal radiation of optical elements and, especially, fluctuations in its intensity during the device operation. The range of measurement of negative temperatures for most modern commercial thermal imagers is limited to −20…−30 °C [4] and does not exceed −40 °C for the best devices [5, 6]. Due to large amount of thermal diagnostic tasks, for example, in cryobiology and cryomedicine, works aimed at creating thermal imaging equipment capable to measure thermal fields of objects with temperatures below –40 °C today is very important.

---

[a] Author to whom correspondence should be addressed: yufomenko@gmail.com

Previously, in collaboration with the scientists from the Argonne National Laboratory (USA), we developed the thermal Imaging System for Laboratory Research (ISLR), the distinctive feature of which was the "open architecture" and the hardware and software modular design [7]. Depending on the research tasks, the device design allows the use of various cooled radiation detectors, lens or mirror objectives, different image scanning format, etc. A distinctive feature of the device design is a special shutter with uniform blackened surface, located at the optical system input as an internal reference source of radiation to improve the accuracy of temperature measurement.

The ISLR prototype for the spectral range 8÷14 μm based on a single-element cooled HgCdTe detector [7] has proven itself to be good for measuring the thermal fields of objects in the temperature range from about −20 °C to +200 °C when solving various scientific tasks [7, 8, 9]. Based on the detector and optical system parameters, we estimated the low-temperature limit of the range of measured temperatures for this imager, which is about −150 °C [10]. However, when using this device to investigate the objects with a temperature about −30 °C and below, we observed a significant deterioration in the accuracy of temperature measurements.

In this paper, in Sec. II we present the limitations of the ISLR imager's low temperature measurement capabilities, as well as an analysis of its optical measurement path and the main sources of temperature measurement errors. In Sec. III we present a new approach to avoid loss of accuracy when measuring the temperature of cold objects. Instead of a completely blackened shutter, we used a shutter with a combined emissivity of its surface. New pixel processing and radiometric calibration are also presented. Finally, to demonstrate the measurement capabilities of our new imager with combined emissivity shutter, in Sec. IV we present a practical example of a low-temperature thermal field studying.

## II. ANALYSIS OF LOW-TEMPERATURE MEASUREMENT CAPABILITIES OF ISLR IMAGER WITH UNIFORMLY BLACKENED SHUTTER

Figure 1 shows the block diagram of ISLR from Ref. [7]. The photo-receiver unit uses a single photoresistor based on $Hg_{0.2}Cd_{0.8}Te$. The detector is installed in a metal optical cryostat and cooled with liquid nitrogen. A germanium lens is used to focus the thermal image in the receiver plane. The photo-receiver unit also includes a low-noise preamplifier, from the output of which the signal is fed to the input of an analog-to-digital converter (ADC). Then the video signal is digitally processed in the electronic unit based on a digital signal processor (DSP).

The IR image raster is formed using a two-coordinate mechanical scanner, which consists of two mirrors (line and frame), which oscillate in mutually perpendicular directions. The mirrors are equipped with sensors showing their position at the given moment so that to unambiguously bind the electrical signal generated by the photo-receiver unit and the position of the instantaneous field of view of the radiation detector in the plane of objects.

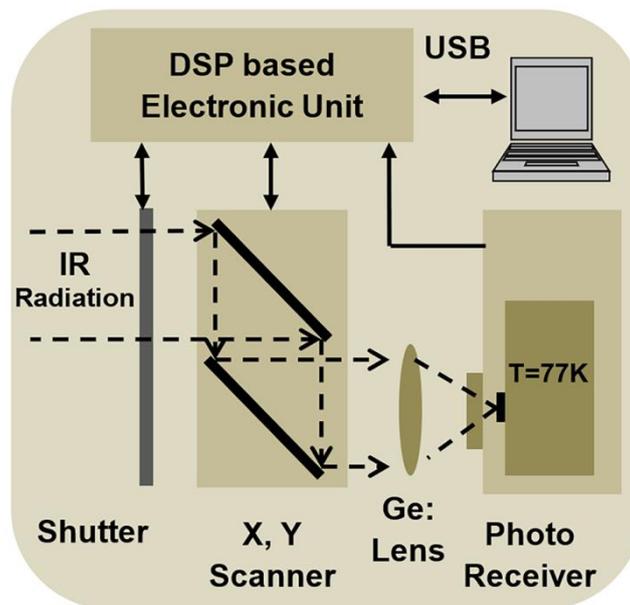

*FIG. 1. Block diagram of thermal Imaging System for Laboratory Research (ISLR).*

The electronic unit generates all the electrical synchronization signals necessary to form an image with a format of 256 × 256 pixels. Each image pixel is represented digitally (12 bits) and carries information about the intensity of thermal radiation of the corresponding area of the observed object. The real-time thermal image with a frame rate of ≈ 1.5 Hz is displayed on a personal computer (PC) monitor. Data are PC-exchanged via a standard USB interface. The OS Windows thermal imager software has a large set of functions for visualization and processing of thermal images, as well as for temperature measuring.

The shutter at the optical system input is made of a material that is opaque to IR radiation and has an emissivity close to 1 due to blackening (see Fig. 2(a)). The shutter temperature is controlled by a semiconductor microthermometer on its surface. The shutter temperature value is digitally available to the imager software and is updated before the start of each image frame. Thus, the shutter can be considered as an internal reference radiation source with controlled temperature.

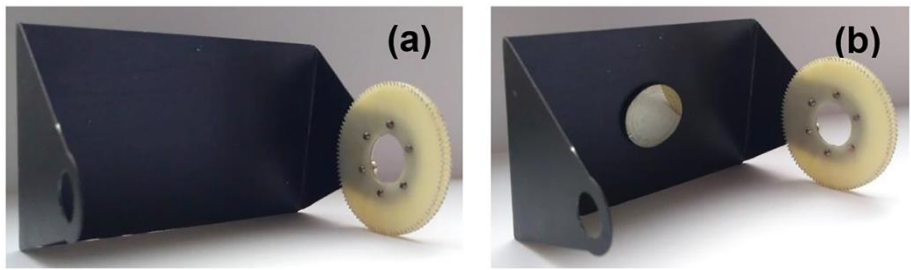

*FIG. 2. Radiation flux shutter: (a) with completely blackened surface, (b) with a combined emissivity surface (to do this, a small mirror was fixed in the center of the shutter that provides efficient reflection of IR radiation in a wide spectral range of wavelengths).*

Prior to start each image frame formation, the shutter covers the entire field of view of the device for a short time, and radiation from its surface hits the detector. At this moment, the value of the detector response voltage $U_{Closed}$ is stored in the RAM of the electronic unit digitally. During the image frame formation, the shutter is open and radiation from the observed object falls on the detector. In this case, the instantaneous value of the signal voltage $U_{X,Y}$ at the output of the photodetector is proportional to the spatial distribution of the radiation intensity on the observed object surface. For each pixel of the image, the processor subtracts from its instantaneous value $U_{X,Y}$ the value of the shutter response signal $U_{Closed}$.

The shutter in the device optical scheme performs important functions. The detector signal in the photo-receiver unit is amplified by an alternating current (see Fig. 3). Subtracting the shutter response signal from the signal for each image pixel restores the DC component of the video signal. The shutter performs the function of a reference radiation source, enabling the measurement of the absolute temperature of the object at any point.

Also, operating with a difference signal makes it possible to exclude the addition of radiation from the optical elements of the device (scanner mirrors, optics, etc.) to the heat flux from the surface of the object being recorded, thereby increasing the accuracy of temperature measurements and their repeatability during operation.

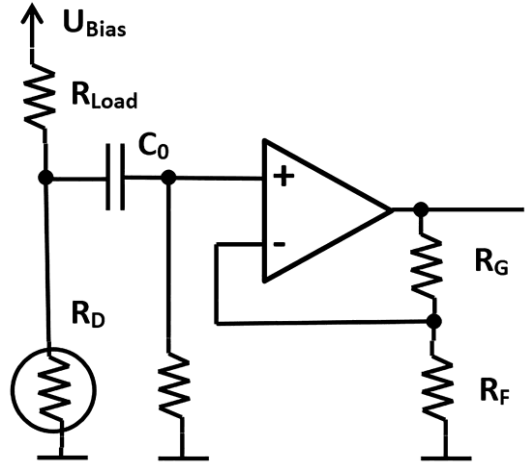

*FIG. 3. Simplified diagram of the preamplifier. Here $U_{Bias}$ is the detector bias voltage, $R_{Load}$ and $R_D$ are the load and detector resistances, respectively, $R_G$ and $R_F$ stand for the feedback resistances, and $C_0$ is a coupling capacitor.*

Indeed, within the time interval when the shutter is closed, the heat flux $P_{Closed}$ that falls on the detector is due to the total contribution of the radiation from the shutter itself $P_{Shut}$ with temperature $T_{Shut}$, radiation from the surface of the scanner mirrors $P_{Mirror}$ and germanium optics $P_{Lens}$, which are at a certain temperature $T_{Int}$, equivalent to the temperature inside the device (see Fig. 4(a)):

$$P_{Closed} = P_{Shut}(T_{Shut}) + P_{Mirror}(T_{Int}) + P_{Lens}(T_{Int}). \qquad (1)$$

Since the response voltage is directly proportional to the power of the thermal radiation fraction absorbed by the detector, we define the response to the heat flux $P_{Closed}$ at the output of the preamplifier as follows:

$$U_{Closed} = SP_{Closed} = U_{Shut} + U_{Mirror} + U_{Lens} - U_{C0}(t), \qquad (2)$$

where $S$ is the detector responsivity in units of V/W, $U_{Shut}$, $U_{Mirror}$, $U_{Lens}$ are the signal voltage components due to the radiation of the shutter, scanner mirrors and optics, respectively, $U_{C0}(t)$ is the voltage drop across the coupling capacitor $C_0$ on the input of the preamplifier, $t$ is time.

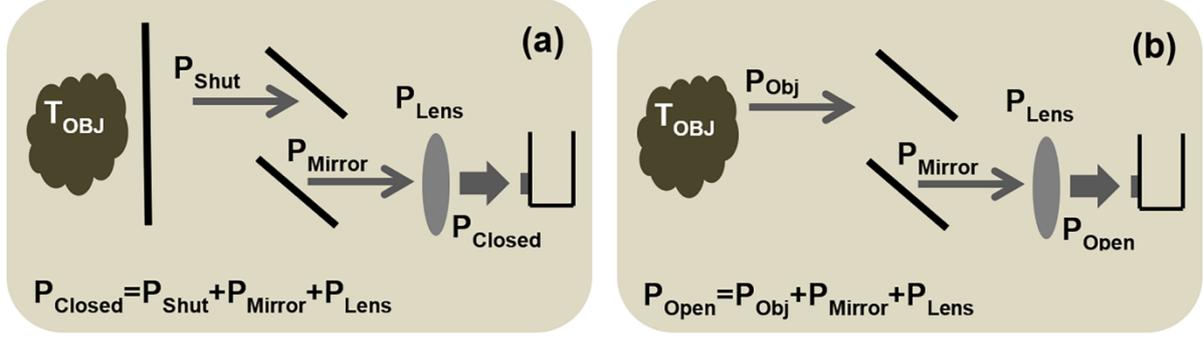

*FIG. 4. Radiation flux at the detector: (a) with closed shutter, (b) with open shutter.*

When the shutter is open, the heat flux $P_{Open}$ falls on the detector, due to the radiation of the surface area of the observed object $P_{ObjX,Y}$ with temperature $T_{ObjX,Y}$ and the radiation of the same optical elements of the device (see Fig. 4(b)):

$$P_{Open} = P_{ObjX,Y}(T_{ObjX,Y}) + P_{Mirror}(T_{Int}) + P_{Lens}(T_{Int}). \tag{3}$$

Accordingly, the signal voltage for any pixel of the image is determined by the expression:

$$U_{X,Y} = SP_{Open} = U_{ObjX,Y} + U_{Mirror} + U_{Lens} - U_{C0}(t), \tag{4}$$

where $U_{ObjX,Y}$ is the component of the detector response to radiation from the studied object for the corresponding image pixel.

Since, as mentioned above, we deal with a differential response signal, the signal voltage $\Delta U_{X,Y}$ for each image pixel is determined by the following expression:

$$\Delta U_{X,Y} = U_{X,Y} - U_{Closed} = U_{ObjX,Y} - U_{Shut}. \tag{5}$$

We believe that the objects observed with the thermal imager are "gray bodies". In this regard, the power of the integral flux of thermal radiation from their surface is determined by the well-known Stefan-Boltzmann law and is proportional to the fourth power of the object temperature [11]. In this case, expression (5) for $\Delta U_{X,Y}$ can be written as follows:

$$\Delta U_{X,Y} = K_1 \delta \varepsilon_{Obj} T^4_{ObjX,Y} - K_2 \delta \varepsilon_{Shut} T^4_{Shut}, \tag{6}$$

where $K_1$ and $K_2$ are the coefficients, $\varepsilon_{Obj}$ and $\varepsilon_{Shut}$ are the radiation coefficients of the object and the shutter, respectively, $\delta = 5.67 \times 10^{-8}$ W/(m$^2$K$^4$) is the Stefan-Boltzmann constant.

The coefficients $K_1$ and $K_2$ take into account the characteristics of the radiation detector, the optical system, as well as all the device design features and generally determined by the expressions:

$$K_1 = \tau_W \tau_{Opt} \varphi S G A_D A_{Lens} / (\pi F^2), \tag{7}$$

$$K_2 = \tau_{Opt}\varphi S G A_D A_{Lens}/(\pi F^2), \tag{8}$$

where $\tau_{Opt}$ is the transmission coefficient of the optics, $\tau_W$ is the transmission coefficient of the input protective window of the thermal imager, $\varphi$ is the fraction of the thermal radiation power in spectral sensitivity range of the detector, $G$ is the electronic path gain, $A_D$ is the radiation detector area, $A_{Lens}$ is the area of the entrance pupil of the lens, $F$ is the focal length of the lens. The coefficients $K_1$ and $K_2$ are slightly different from each other, since the device has an input protective window, providing an additional absorption for the heat flow from the observed object.

Assuming that the shutter emissivity is close to 1, the temperature of the object at any point in thermal image can be calculated from the expression, which follows from Eq. (6):

$$T_{ObjX,Y} = \sqrt[4]{\frac{\frac{\Delta U_{X,Y}}{\delta} + K_2 T_{Shut}^4}{\varepsilon_{Obj} K_1}}. \tag{9}$$

As it was demonstrated in Eq. (9), the advantages of using a shutter element in the device optical scheme are obvious. First of all, the effect of thermal radiation from the components of the optical system and fluctuations in its intensity on the accuracy of temperature measurement is excluded. Such fluctuations inevitably arise during their exploitation, when environmental conditions change, etc. It should be noted that in most cases modern thermal imagers, having the aim to achieve compactness and simplify the design, do not have such mechanical elements. To compensate for the influence of optical radiation, they use special algorithms that require a large number of additional calibrations [12].

Equation (9) is directly used by the thermal imager software to calculate the absolute temperature in an arbitrary pixel of a thermal image. The input data for the calculation are the signal voltage $\Delta U_{X,Y}$ for the corresponding image pixel, the shutter temperature $T_{Shut}$ at the moment of shooting, and the emissivity of the object $\varepsilon_{obj}$. The temperature of the shutter is quite accurately controlled by a microthermometer. In this regard, changes in its temperature (drift or sharp fluctuations when environmental conditions change) do not affect the results of temperature measurements. To calculate the $K_1$ and $K_2$ coefficients, a reference radiation source with a known temperature (blackbody model) is used. $K_1$ and $K_2$ are calculated from the inverse expression of Eq. (9) for two temperature points of the blackbody.

The IR detector is cooled with liquid nitrogen and its responsivity remains stable during the refrigerant evaporation. Temperature fluctuations in the environment do not lead to significant changes in the transmission coefficient of the electronic amplifying path and their influence on the degradation of the coefficients $K_1$ and $K_2$ can also be neglected.

Figure 5 (curve A) shows the temperature measurement error $\Delta T_{ObjX,Y}$ depending on the temperature of the registered object in the range −130…+30 °C. By error we mean the difference between the value of the temperature of the object $T_{ObjX,Y}$, measured by our thermal imager, and the value of its true temperature, measured with the help of a metrologically verified contact thermometer. In the range from −30 °C to +30 °C the measurements were carried out with a Fluke Portable Infrared Calibrator-9133 blackbody metrology sources (Fluke Corporation, USA). To obtain lower temperatures, down to −150 °C, we used homemade model of a blackbody cooled by liquid nitrogen. The blackbody surface temperature was controlled by a thermocouple. Coefficients $K_1$ and $K_2$ were calculated at the temperature points of 0 °C and −10 °C, and their values for the current design were 0.793 m$^2$V/W and 0.810 m$^2$V/W, respectively.

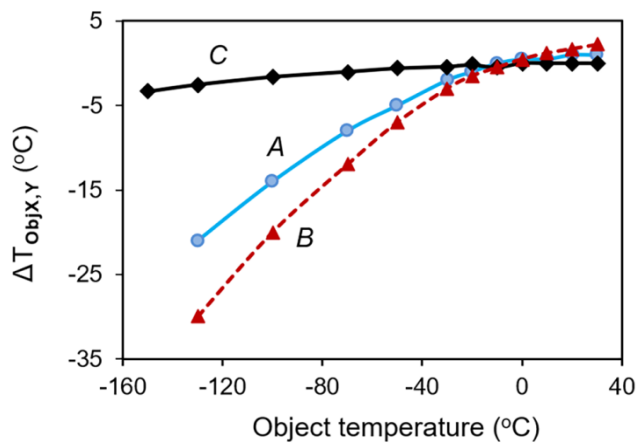

*FIG. 5. Temperature measurement error $\Delta T_{ObjX,Y}$ vs temperature of the registered object in the range −130…+30 °C: A - when using a completely blackened shutter, B - calculated from Eq. (10), C - when using a shutter with a combined emissivity.*

As Figure 5 demonstrates (see curve A), the error increases significantly in the low-temperature region and is systematic, not random. We attributed this behavior of error to the methodological error of indirect measurements made by this thermal imager.

Since the photo-receiver unit uses a semiconductor photoresistor with a sensitivity in a narrow spectral range of 8÷14 μm, only a part of the integral heat flux emitted by the surface of objects in the observed scene is effectively converted into an electrical signal. The power of this part of the heat flux is determined not only by the temperature of the objects according to the Stefan-Boltzmann law, but also by the fraction $\varphi$ of the thermal radiation power coming on the spectral sensitivity range of the detector, and, in turn, also depends on the temperature of the object in accordance with the Wien's displacement law.

Figure 6 shows the fraction of the thermal radiation power of the blackbody, coming in the spectral range of 8÷14 μm, depending on the temperature of the blackbody (according to Ref. [10]). Obviously, resulted from the temperature dependence of $\varphi$, in order to accurately calculate the temperature from expression (9), the coefficient $K_1$ must also change depending on the temperature of the observed object. This follows from expression (7) for $K_1$, where $\varphi$ enters as a factor. However, the absolute temperature calculation algorithm used in the thermal imager does not take into account the temperature dependence of $\varphi$. For the entire range of measured temperatures a constant value $\varphi_0$ = const is used, corresponding to the blackbody temperature at which the calibration was performed. Since the temperatures of the observed objects can differ significantly from the calibration temperature, their calculation according to Eq. (9) will give an inaccurate result. In the region of calibration temperature points (0 °C and −10 °C), the accuracy will be the best, since $\varphi_0$ corresponds to the temperature of the observed object. In the range of measured temperatures about −20…+200 °C, the spectral fraction $\varphi$ changes quite weakly. This does not lead to significant changes in the accuracy of temperature measurements. However, for temperatures much lower than about −20 °C, the spectral fraction $\varphi$ changes significantly. Therefore, when the temperature of the object under study decreases, the applied temperature calculation algorithm leads to a noticeable deterioration in the measurement accuracy.

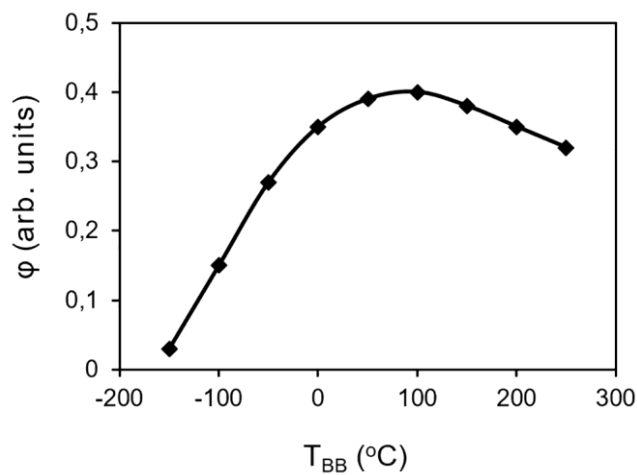

*FIG. 6. Fraction of blackbody thermal radiation power $\varphi$ in the spectral range 8÷14 μm depending on the blackbody temperature $T_{BB}$.*

The coefficient $K_2$ can be considered constant, since the temperature of the shutter during the operation of the thermal imager changes insignificantly in relation to the temperature $T_{Shut}$ at which the calibration was performed.

Using the differential method for finding indirect measurement errors [13], from expression (9) we can obtain an estimated relation that determines the temperature measurement error $\Delta T_{ObjX,Y}$ depending on the error of the coefficient $K_1$ due to the temperature dependence of the spectral fraction $\varphi$:

$$\Delta T_{ObjX,Y} = -\frac{1}{4} T_{ObjX,Y} \frac{\Delta \varphi}{\varphi_0}, \tag{10}$$

$$\Delta \varphi = \varphi_0 - \varphi_{ObjX,Y}, \tag{11}$$

where $\varphi_0$ is the spectral fraction of the radiation power corresponding to the temperature at which the calibration was performed, $\varphi_{ObjX,Y}$ is the spectral fraction of the radiation power corresponding to the temperature of the object.

Expression (10) is approximate and is valid for small increments $\Delta \varphi$. However, the character of this dependence (see curve $B$ in Fig. 5) allows us to conclude that it is the neglect of the temperature dependence of the coefficient $K_1$ that can be the main source of measurement error. It should be noted that it is very difficult to take into account the temperature dependence of $K_1$ for a more accurate calculation of $T_{ObjX,Y}$ using Eq. (9).

## III. COMBINED EMISSIVITY SHUTTER IN THERMAL IMAGING CAMERA DESIGN TO DIMINISH LOW-TEMPERATURE MEASUREMENT ERROR

For measuring temperature with thermal imager, it is a common practice to use the calibration characteristic $T_{BB}=f(U_{BB})$ measured at discrete temperature points using a temperature-controlled blackbody model, where $T_{BB}$ is the blackbody temperature and $U_{BB}$ is the signal voltage. However, it is impossible to use $\Delta U_{X,Y}$ to build a calibration characteristic and accurately determine the temperature of the objects under observation in our case. Despite the fact that when using a differential signal, the response components due to the radiation of the optical elements of the device are completely excluded, the resulting voltage $\Delta U_{X,Y}$ is determined by the difference in signals to heat fluxes from the surface of the object and the shutter. Since the temperature of the shutter depends on the operating conditions of the device and can change with time over a fairly wide range, it is not possible to neglect the contribution of its heat flux without losing the accuracy of temperature measurements.

On the other hand, by determining in one way or another the detector signal $U_{Shut}$ to the heat flux from the shutter surface, one can find the voltage value $U_{ObjX,Y}$, which is determined by the intensity of thermal radiation only from the observed object. Conforming to the expression (5):

$$U_{ObjX,Y} = \Delta U_{X,Y} + U_{Shut}. \tag{12}$$

We improved the device optical path, which made it possible to determine the $U_{Shut}$ component before the formation of each frame of a thermal image. Instead of a uniform blackened shutter at the input of the optical system, we used a shutter with a combined emissivity (see Fig. 2(b)). To do this, a small mirror was fixed in the center on its surface that provides efficient reflection of IR radiation in a wide spectral range of wavelengths (reflection coefficient $\gamma \approx 0.95$). The mirror is a thin glass plate, on the polished surface of which a layer of aluminum is deposited.

Before the formation of each image frame, as discussed above, the shutter for a short time completely covers the entire field of view of the device. When a section of the shutter surface with a mirror enters the field of view of the photo-receiver unit (this position is precisely fixed due to the sensors of the instantaneous position of the scanner), the radiation flux, created by the detector itself and its cooled aperture diaphragm is reflected from the mirror and falls back on the detector. Since the detector is cooled with liquid nitrogen, such a mirror section of the shutter can be considered as a very cold radiation source. Of course, owing to the thermal radiation of the mirror itself, which is at room temperature, its radiation temperature exceeds the boiling point of liquid nitrogen.

When the shutter is closed and a mirror section enters the detector's field of view, the heat flux $P^0_{Closed}$ in the receiving plane consists of the radiation of the detector $P_D$ reflected from the mirror, which is at a temperature of 77 K, the radiation $P^0_{Shut}$ of the mirror itself on the shutter surface with temperature $T_{Shut}$ and the radiation of optical elements of the device with temperature $T_{Int}$. We estimate the power of the thermal radiation of the detector with a temperature of 77 K reflected from the mirror to be $P_D \approx 10^{-12}$ W. At the same time, the radiation power of the mirror itself, which has a room temperature of 300 K, is at the level of $P^0_{Shut} \approx 10^{-9}$ W. Since $P_D$ is much smaller than $P^0_{Shut}$, the contribution of the reflected detector radiation to the total heat flux can be neglected. Thus, we obtain:

$$P^0_{Closed} = P_{Mirror}(T_{Int}) + P_{Lens}(T_{Int}) + P^0_{Shut}(T_{Shut}). \tag{13}$$

Accordingly, when the mirror section of the shutter enters the field of view of the detector, the response voltage is determined by the expression:

$$U_{Closed}^0 = SP_{Closed}^0 = U_{Mirror} + U_{Lens} + U_{Shut}^0 - U_{C0}(t), \quad (14)$$

where $U^0_{Shut}$ is the signal component due to the mirror radiation on the shutter surface.

When the shutter is closed and its blackened area enters the field of view of the detector, the heat flux $P_{Closed}$ and the response voltage $U_{Closed}$ are due to the radiation of the optical elements of the device and the blackened shutter Eq. (1) and Eq. (2), respectively. Values $U_{Closed}$ and $U^0_{Closed}$, corresponding to the moments of time when the blackened and mirror areas of the shutter enter the field of view, are stored in the RAM of the electronic unit before the formation of each frame of the thermal image.

The differential response voltage $\Delta U_{Closed}$ to heat fluxes from the surface of the blackened and mirror areas of the shutter in accordance with expressions (2) and (14) is equal to:

$$\Delta U_{Closed} = U_{Closed} - U_{Closed}^0 = U_{Shut} - U_{Shut}^0. \quad (15)$$

Since the shutter temperature is the same for both areas, the difference in the intensity of heat fluxes from their surfaces is stipulated only by the difference in emissivity. Accordingly, assuming that the emissivity of the blackened area of the shutter is close to 1, the values of $U_{Shut}$ and $U^0_{Shut}$ are related by the expression:

$$U_{Shut}^0 = (1-\gamma)U_{Shut}, \quad (16)$$

where $\gamma$ is the reflection coefficient of the mirror area of the shutter. In this case, from expression (15) we obtain:

$$U_{Shut} = \frac{1}{\gamma}\Delta U_{Closed}. \quad (17)$$

Now, using expressions (12) and (17), for each pixel of the image we can determine the signal voltage, due solely to the heat flux from the registered object:

$$U_{ObjX,Y} = \Delta U_{X,Y} + \frac{1}{\gamma}\Delta U_{Closed}. \quad (18)$$

Finally, having deciphered $\Delta U_{X,Y}$ and $\Delta U_{Closed}$ in expression (18) and assuming $\gamma \approx 0.95$, we determined the signal voltage for each image pixel from the following expression:

$$U_{ObjX,Y} = (U_{X,Y} - U_{Closed}) + 1.053(U_{Closed} - U_{Closed}^0). \quad (19)$$

Since $U_{Closed}$ and $U^0_{Closed}$ are updated before each image frame starts, there is no need to control the temperature of the shutter to make temperature measurements. It is also not necessary to take into account the shutter temperature when calibrating the device.

Using a blackbody with adjusted temperature, we measured the detailed calibration characteristic $U_{BB}=f(T_{BB})$ at discrete temperature points in the range −150...+200 °C, where $U_{BB}$ is the signal voltage, which was calculated from expression (19). For the calculation, the averaged value $U_{X,Y}$ of the pixel array in the center of the blackbody thermal image was used. To carry out temperature measurements, we approximated the inverse calibration characteristic $T_{BB}=f(U_{BB})$ (see Fig. 7) with a cubic polynomial dependence of the following form:

$$T_{BB}(U_{BB}) = a_0 U_{BB}^3 + a_1 U_{BB}^2 + a_2 U_{BB} + a_3, \quad (20)$$

where $a_0, a_1, a_2, a_3$ are the coefficients of the polynomial. Using the approximation of the calibration characteristic (20), the so-called radiation temperature $T_{BB}(U_{ObjX,Y})$ for any pixel of the thermal image is determined. The radiation temperature of an object (a "gray" body with emissivity < 1) is the temperature of a blackbody at which its radiant flux power is equal to the radiant flux power from the object. To determine actual temperature for any pixel (in units of °C), we used the equation, which follows directly from the Stefan-Boltzmann law:

$$T_{ObjX,Y} = \frac{T_{BB}(U_{ObjX,Y})}{\sqrt[4]{\varepsilon_{Obj}}} - 273.15. \quad (21)$$

In Fig. 5 the curve $C$ is the experimental dependence of an error when the object temperature was obtained from expression (21). Since we were interested in improving the characteristics of the thermal imager for studying low-temperature thermal fields, the measurements were carried out in the temperature range of the studied objects −150...+30 °C.

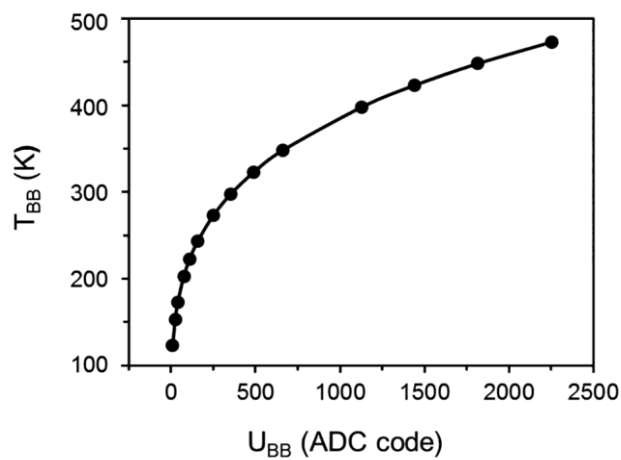

Fig. 7. Inverse calibration characteristic $T_{BB}=f(U_{BB})$. The signal $U_{BB}$ (in ADC code) for each blackbody temperature $T_{BB}$ was determined from Eq. (19).

Obviously, the approach we proposed enables the achieving of a proper measurement accuracy in the low-temperature region. The relative measurement error does not exceed 3% (object temperature $T_{Obj} = -150$ °C) and is associated with a deterioration in the temperature resolution of the receiving and electronic paths with a decrease in the temperature of the recorded objects (as a result of a significant decrease in the intensity of thermal radiation in the spectral range of the detector sensitivity). An additional contribution is made by the error associated with the limited resolution of the digitization system, which is determined by the least significant bit of the ADC.

It should be noted that a fairly significant parameter on which the results of thermal measurements depend is the emissivity $\varepsilon_{Obj}$. All thermal imagers and single-band pyrometers, regardless of their design and measurement algorithm, use the emissivity value as an input parameter to calculate the temperature of the object under study. Often, in real conditions, it is impossible to measure emissivity with acceptable accuracy for one reason or another. Therefore, in practice, the estimated value is used, which is the emissivity value taken from the reference and which is measured under certain conditions (temperature, spectral range, surface condition, etc.). From the point of view of thermal imaging, most materials are "gray" bodies with an emissivity less than 1, which is weakly dependent on temperature. However, a number of materials, so-called "non-gray" bodies (for example, metals), have a noticeable temperature dependence of emissivity. In this case, the estimated and actual emissivity values may differ from each other, which leads to an additional error in temperature measurements (effective emissivity estimation error). Surface condition (wet or dry, clean or dirty, etc.) can also be a source of estimation emissivity error. In some cases, this additional error can even exceed the main measurement error of the thermal imager. To study the temperature dependence of the measurement error, we used a blackbody model, the emissivity of which is close to 1 and, by definition, does not depend on the temperature of the blackbody and the radiation wavelength. In this regard, our results (see curve C in Fig. 5) demonstrate the best accuracy that can be achieved in the device discussed in this paper as a result of improvements in design and pixel processing. In real conditions, due to the difference between the estimated and actual values of emissivity, a deterioration in the accuracy of measuring the temperature fields of objects with poorly studied properties of the radiating surface may occur. This kind of additional error is inherent in all thermal imagers, regardless of their design, and its contribution can only be minimized by accurately determining the emissivity of the surface under study.

## IV. MONITORING OF LOW-TEMPERATURE THERMAL FIELD DYNAMICS

The upgraded thermal imager, described above, was successfully used for remote monitoring of the thermal field dynamics during freeze-thawing of biological tissues in vivo [14]. Figure 8(a) shows a simplified diagram of the experimental setup: 1 – experimental animal, 2 – cryoapplicator cooled with liquid nitrogen, 3 – the thermal imager. The cryoapplicator was applied to the skin area of the experimental animal for 1.5 minutes. After the end of cryoexposure, we used a thermal imager to monitor the area where the applicator was exposed, on which an ice spot had formed, and recorded a thermographic film. Figure 8(b) shows a thermal image of an ice spot on the animal's skin taken 2 seconds after the end of cryoexposure. Figure 8(c) shows the temperature distribution along the diameter of the ice spot (i. e. along the white arrow in Fig. 8(b)) imaged 2, 20, and 600 seconds after the end of cryoexposure.

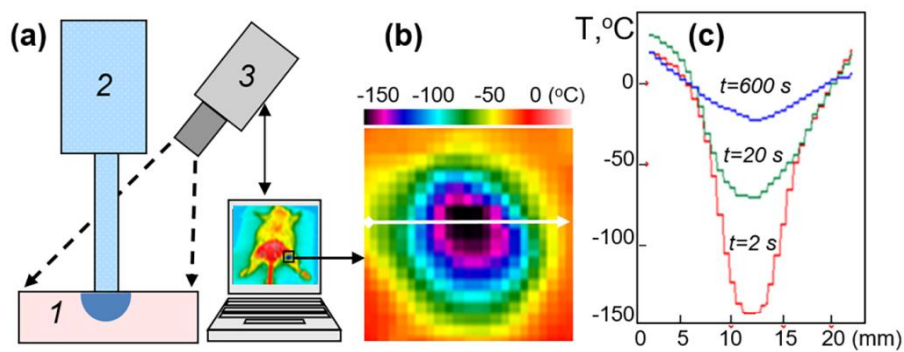

*FIG. 8. Using the thermal imager to monitor freezing-thawing of biological tissues in vivo: (a) simplified diagram of the experimental setup (1 – experimental animal, 2 – cryoapplicator cooled with liquid nitrogen, 3 – thermal imager), (b) a thermal image of the ice spot caused by cryoexposure of 1.5 minutes, (c) temperature distribution along the diameter of the ice spot 2, 20, and 600 seconds after the end of cryoexposure.*

## V. CONCLUSION

To summarize, a thermal imager based on a single-element IR detector, cooled with liquid nitrogen, with a spectral sensitivity range of 8÷14 μm was developed to measure low-temperature thermal fields down to −150 °C. To measure the temperature of the objects under study, a special shutter with a combined emissivity was used as an internal reference source of radiation in the optical scheme of the device. For this purpose a small area with a high reflection coefficient of IR radiation was formed on the surface of the blackened shutter. The use of a differential detector response signal to the shutter blackened and mirror areas for processing each pixel of a thermal image made it possible to significantly improve the accuracy of measuring the temperature of cold objects. The device can be used in studies of various low-temperature processes, including measurements of thermal field dynamics during freeze-thawing of biological tissues *in vivo*.

## ACKNOWLEDGMENTS

The authors acknowledge useful discussion with O.G. Turutanov and technical assistance of A.I. Ryzhov and M.P. Liul. This work was supported by the National Research Foundation of Ukraine by the grant titled "Thermal imaging study of soft tissues with thermal injury and mathematical modeling of the processes accompanying it" (Grant No. 2022.01/0094, Grant State Registration No. 0123U103506).

## AUTHOR DECLARATIONS

### Conflict of Interest

The authors have no conflicts to disclose.

### Author Contributions

**Eduard Gordiyenko**: Conceptualization (lead); Data curation (lead); Formal analysis (lead); Investigation (lead); Methodology (lead); Validation (equal); Visualization (lead); Writing – original draft (lead); Writing – review & editing (lead). **Yulia Fomenko**: Data curation (equal); Formal analysis (equal); Validation (equal); Writing – review & editing (equal). **Galyna Shustakova**: Data curation (equal); Investigation (equal); Methodology (equal); Visualization (equal); Writing – review & editing (equal). **Gennadiy Kovalov**: Data curation (equal); Investigation (equal); Methodology (equal); Resources (supporting); Writing – review & editing (equal). **Sergiy Shevchenko**: Conceptualization (equal); Funding acquisition (lead); Supervision (lead); Writing – review & editing (equal).

## DATA AVAILABILITY

The data that support the findings of this study are available from the corresponding author upon reasonable request.

**REFERENCES**

1. Glenn J. Tattersall, "Infrared thermography: a non-invasive window into thermal physiology," Comp. Biochem. Physiol. Part A 202, 78−98 (2016). https://doi.org/10.1016/j.cbpa.2016.02.022

2. P. Pasquali (ed.), *Cryosurgery. A practical manual* (Springer−Verlag, Berlin, Heidelberg, 2015).

3. E.E. Zimmerman, P. Crawford, "Cutaneous cryosurgery," Am. Fam. Physician 86(12), 1118 (2012). Available at https://pubmed.ncbi.nlm.nih.gov/23316984/

4. A.N. Wilson, Khushi Anil Gupta, Balu Harshavardan Koduru, Abhinav Kunar, Ajit Jha, and Linga Reddy Cenkeramaddi, "Recent advances in thermal imaging and its applications using machine learning: a review," IEEE Sensors Journal 23(4), 3395−3407 (2023). https://doi.org/10.1109/JSEN.2023.3234335

5. *Product catalogue 2018* (FLIR Systems Inc., USA, 2018); available at http://www.flir.kiev.ua/pdf/flir-product-catalog.pdf ; accessed 20 October 2023.

6. *Product guide 2023-2024* (FLUKE Corporation, 2023); available at https://s7d1.scene7.com/is/content/fluke/fluke/en-us/2022/marketing/flk-220396-en-2023-2024-product-guide/final-files/220396-en-2023-2024-product-guide-complete.pdf ; accessed 20 October 2023.

7. V. Yefremenko, E. Gordiyenko, G. Shustakova, Yu. Fomenko, A. Datesman, G. Wang, J. Pearson, E. E. W. Cohen, and V. Novosad, "A broadband imaging system for research applications," Rev. Sci. Instrum. 80, 056104 (2009). https://doi.org/10.1063/1.3124796

8. E. E. W. Cohen, O. Ahmed, M. Kocherginsky, G. Shustakova, E. Kistner-Griffin, J. K. Salama, V. Yefremenko, and V. Novosad, "Study of functional infrared imaging for early detection of mucositis in locally advanced head and neck cancer treated with chemoradiotherapy," Oral Oncol. 49(10), 1025−1031 (2013). https://doi.org/10.1016/j.oraloncology.2013.07.009

9. E. Gordiyenko, N. Glushchuk, Yu. Fomenko, G. Shustakova, I. Dzeshulskaya, and Y. Ivanko, "Nondestructive testing of composite materials of aircraft elements by active thermography," Sci. Innov. 14(2), 37−45 (2018). https://doi.org/10.15407/scine14.02.037

10. R. Siegel, and J. Howell, *Thermal radiation heat transfer, 4-th edition* (Taylor & Francis, New York, London, 2002).

11. K. Chrzanowski, *Non-contact thermometry: measurement errors* (Polish Chapter of SPIE, Warsaw, 2001).

12. A. Tempelhahn, H. Budzier, V. Krause, and G. Gerlach, "Shutter-less calibration of uncooled infrared cameras," J. Sens. Sens. Syst. 5, 9−16 (2016). https://doi.org/10.5194/jsss-5-9-2016

13. M. Sedlacek, V. Haasz, *Electrical measurements and instrumentations* (Vidavatelstvo CVUT, Prague, 1996).

14. G. Kovalov, G. Shustakova, E. Gordiyenko, Yu. Fomenko, and M. Glushchuk, "Infrared thermal imaging controls freezing and warming in skin cryoablation," Cryobiology 103, 32 (2021). https://doi.org/10.1016/j.cryobiol.2021.09.014